# Lattice Softening in Metastable bcc Co$_x$Mn$_{100-x}$(001) Ferromagnetic Layers for a Strain-Less Magnetic Tunnel Junction

Kelvin Elphick,[1] Kenta Yoshida,[2] Tufan Roy,[3] Tomohiro Ichinose,[4] Kazuma Kunimatsu,[4,5] Tomoki Tsuchiya,[6,7] Masahito Tsujikawa,[3,7] Yasuyoshi Nagai,[2] Shigemi Mizukami,[4,6,7] Masafumi Shirai [3,6,7] and Atsufumi Hirohata [1]

[1] *Department of Electronic Engineering, University of York, Heslington, York YO10 5DD, United Kingdom*

[2] *Institute for Materials Research, Tohoku University, Katahira 2-1-1, Sendai 980-8577, Japan*

[3] *Research Institute of Electrical Communication, Tohoku University, Sendai 980-8577, Japan*

[4] *WPI Advanced Institute for Materials Research, Tohoku University, Katahira 2-1-1, Sendai 980-8577, Japan*

[5] *Department of Applied Physics, Graduate School of Engineering, Tohoku University, Sendai 980-8579, Japan*

[6] *Center for Science and Innovation in Spintronics (CSIS), Core Research Cluster (CRC), Tohoku University, Sendai 980-8577, Japan*

[7] *Center for Spintronics Research Network (CSRN), Tohoku University, Sendai 980-8577, Japan*

**Abstract**

In spintronics, one of the long standing questions is why the MgO-based magnetic tunnel junction (MTJ) is almost the only option to achieve a large tunnelling magnetoresistance (TMR) ratio at room temperature (RT) but not as large as the theoretical prediction. This study focuses on the development of an almost strain-free MTJ using metastable bcc Co$_x$Mn$_{100-x}$ ferromagnetic films. We have investigated the degree of crystallisation in MTJ consisting of Co$_x$Mn$_{100-x}$/MgO/Co$_x$Mn$_{100-x}$ ($x$ = 66, 75, 83 and 86) in relation to their TMR ratios. Cross-sectional high resolution transmission electron microscopy (HRTEM) reveals that almost consistent lattice constants of these layers for 66 ≤ $x$ ≤ 83 with maintaining large TMR ratios of 229% at RT, confirming the soft nature of the Co$_x$Mn$_{100-x}$ layer with some dislocations at the MgO/Co$_{75}$Mn$_{25}$ interfaces. For $x$ = 86, on the other hand, the TMR ratio is found to be reduced to 142% at RT, which is partially attributed to the increased number of the dislocations at the MgO/Co$_{86}$Mn$_{14}$ interfaces and amorphous grains identified in the MgO barrier. *Ab-initio* calculations confirm the crystalline deformation stability across a broad compositional range in CoMn, proving the advantage of a strain-free interface for much larger TMR ratios.

**Introduction**

A magnetic tunnel junction (MTJ) is known as a crucial spintronic device because of its broad applications in read heads for hard disk drives (HDDs), magnetic random access memories (MRAMs) and magnetic sensors. This has incurred extensive studies in many research institutions worldwide [1][2][3]. The most commonly used MTJs are composed with CoFeB ferromagnetic electrodes with a MgO tunnelling barrier. High tunnelling magnetoresistance (TMR) ratios of > 600% were obtained at room temperature (RT) [4][5][6][7]. Recently, the other ferromagnetic alloys such as half-metallic ferromagnetic Heusler alloys and metastable ferromagnets, *e.g.*, $Co_{75}Mn_{25}$, have been employed as electrodes in MTJ [8][9][10]. However, the corresponding TMR ratios measured experimentally are still far below from theoretical predictions of > 1,000% at RT [11][12].

Numerous methods were proposed to improve the degree of crystallisation in MTJ. One of the most commonly used methods for MTJ fabrication is a thermal annealing process under a magnetic field. It has been reported that for an as-deposited MTJ, the excessive oxygen content in the MgO barrier is accumulated at the interface near the top electrode. After the annealing process, the excessive oxygen is evenly distributed in the barrier [13][14]. The annealing process also plays an important role in the MTJ crystallisation. The crystallinity of the MgO barrier and the ferromagnetic electrodes facilitates coherent tunnelling [7]. For instance, a commonly used ferromagnetic electrode such as CoFeB is crystallised via the MgO crystallisation after the annealing process [15]. It has been reported that without annealing, the MgO barrier exhibits an amorphous phase within a region in a few nanometres above the bottom electrode [16].

The crystallisation degree of the MTJ can be determined by measuring the saturation magnetisation and crystallographic structure [17][18]. Cross-sectional high resolution transmission electron microscopy (HRTEM) images for MTJ using different annealing conditions have been reported to date [4][19]. An epitaxial crystallisation of the entire MgO layer has been obtained under the annealing condition of 90 minutes at 420°C [20]. Amorphous and polycrystalline grains were observed when MTJ was annealed for 3 minutes instead [20]. However, excessive duration of annealing and annealing at higher temperatures cause diffusion of constituent elements in MTJ, namely Ru, Mn and B [4][14][21]. Butler *et al.* have reported that interfacial roughness and bonding at the electrode/insulator interface are the critical factors to determine the corresponding spin-polarised electron transport and resulting TMR ratios [10].

However, the detailed mechanism of the crystallisation process and the corresponding interfacial strain as well as their resulting advantages of MgO over the other epitaxial barriers have yet to be understood conclusively. Recently we reported a high TMR ratio over 600% (200%) at 10K (RT) in metastable bcc $Co_xMn_{100-x}$ (CoMn)/MgO/CoMn MTJs and discussed the relationship between their magnetism and TMR effect [22][23]. However, the relationship between their TMR ratios and their crystalline structures is not yet clear. In this paper, we provide atomic analysis on MTJ consisting of $Co_xMn_{100-x}$ (CoMn)/MgO/CoMn, including their degrees of crystallisation. Cross-sectional HRTEM imaging was carried out to understand the atomic structures at the CoMn/MgO/CoMn interfaces and within the MgO barrier. Inhomogeneous crystallisation was observed especially within the MgO barrier depending on the bottom CoMn crystallinity after the annealing process. The lattice constants estimated from these images were compared with *ab-initio* calculations to confirm the softness of CoMn. Based on this approach, we explain why the TMR ratios experimentally reported are smaller than the theoretically expected values and why MgO barriers are advantageous for the demonstration of large TMR ratios.

**Experimental Procedures**

MTJ samples were prepared using conventional magnetron sputtering with a base pressure of $2 \times 10^{-7}$ Pa. MgO(001) single crystal substrates were used for the deposition of MTJ consisting of Cr (40)/$Co_xMn_{100-x}$ (10)/MgO (2.4 and 20)/$Co_xMn_{100-x}$ (4)/$Co_3Fe$(1.5)/IrMn (10)/Ru (5) (thickness in nm) with a Ti and Au top electrode as illustrated in the inset of Fig. 1(a) [22][23]. Two different thicknesses of the MgO layers were used for MTJs with 2.4 nm and reference samples with 20 nm. For CoMn, four different nominal compositions of $x$ = 66, 75, 83 and 86 were employed, allowing to control the lattice constants of the bottom and top electrodes precisely [23]. The crystalline phase of $Co_xMn_{100-x}$ in bulk becomes face-centred cubic (fcc)/hexagonal closed packing (hcp) for those range of $x$ [24]. However, it has been reported that ferromagnetic $Co_xMn_{100-x}$, body-centred cubic (bcc) phase can be obtained around $x$ = 75 in epitaxial films [25]. In order to achieve these crystalline engineering in MTJ, *in-situ* annealing was carried out after the deposition of the Cr seed layer and $Co_xMn_{100-x}$ layers at 700°C and 200°C, respectively. MTJs were then post-annealed at 325°C for their crystallisation after patterned into pillars with Ti/Au electrodes by photolithography [23].

TMR measurements were carried out using the conventional four-terminal setup with elevating temperature and the corresponding atomic structures at the CoMn/MgO/CoMn interfaces were studied to understand the compositional dependence of the $Co_xMn_{100-x}$ MTJ samples. This study primarily focuses on two samples which are $x$ = 75 and 86 due to their distinctive TMR ratios of 229% and 142%, respectively. Cross-sectional TEM specimens were prepared using a mechanical polishing technique. HRTEM images were taken using JEOL JEM-2100 Plus at 600k magnification. These results were compared with *ab initio* calculations on the lattice stability of the $Co_xMn_{100-x}$ alloys.

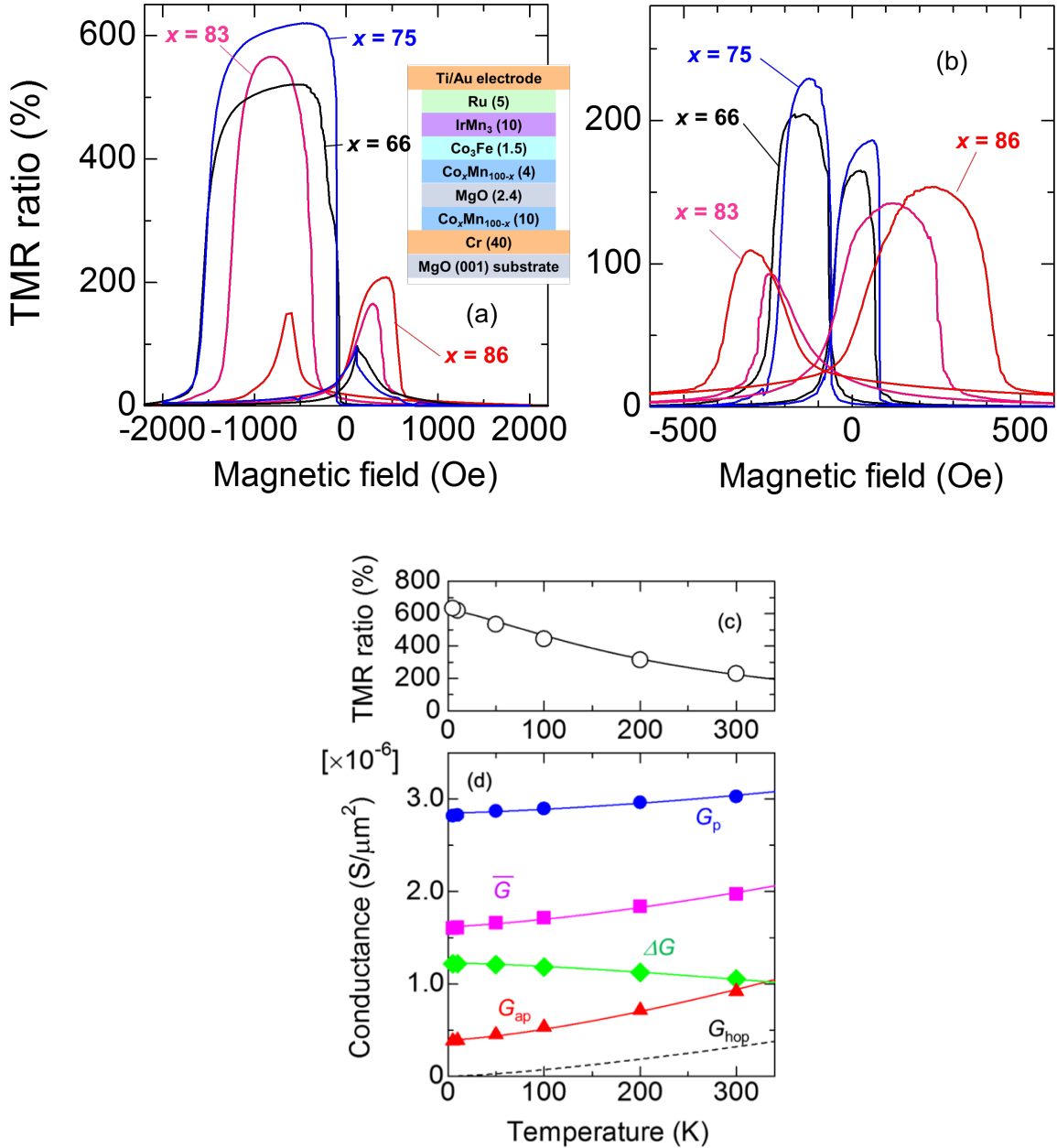

Figure 1 TMR curves of $Co_xMn_{100-x}$ (66 ≤ $x$ ≤86) measured at (a) 10 K and RT (300 K). The inset shows a schematic diagram of MTJ structures studied [22][23]. (c) Temperature dependence of the TMR ratios for MTJ with $x$ = 75 [22]. (d) Temperature dependence of the conductance for parallel state $G_p$ and antiparallel state $G_{ap}$ as well as the spin-dependent and -independent parts of the conductance, $\bar{G}$ and $\Delta G$. $G_{hop}$ is the spin-independent conductance due to the two-step hopping via defect states in a MgO barrier.

**Results and Discussion**

First TMR ratios of $Co_xMn_{100-x}$-based MTJs were measured between 4 and 300 K as shown in Fig. 1. The TMR ratios measured at 10 K and 300 K show almost square curves for the lower Co concentration of *x* = 75 and 66 as seen in Figs. 1(a) and 1(b) [22][23], while the curves for the higher Co concentration of *x* = 86 and 83 at 10K and 300K contain stronger components of magnetisation rotation [see Figs. 1(a) and 1(b)]. This suggests incomplete antiparallel states of magnetisations for *x* = 86 and 83, which may be partially due to the large coercivity for the 10nm thick bottom CoMn layer with *x* = 86 and 83 and also the presence of magnetic volume with spin fluctuation at the interfaces. The difference between *x* = 86 and 83, and 75 and 66 may be related to the stability of the metastable bcc phase against the post-annealing at 325°C. The former MTJs with the higher Co concentration are less stable than the latter MTJs in this thickness [23]. The temperature dependence of the TMR ratios for *x* = 75 is shown in Fig. 1(c), revealing the decreases in the TMR ratios with increasing temperature. In the magnetic tunnelling conductance for *x* = 75 show in Fig. 1(d), both parallel and antiparallel conductance, $G_p$ and $G_{ap}$, respectively, are found to increase with increasing temperature. These data are analysed using Shang's model [26]:

$$G_{\text{p (ap)}} = G_0(T)[1 + (-)P_0^2 m^2(T)] + G_{\text{hop}}(T). \tag{1}$$

Here, $G_0$ is the mean conductance, $P_0$ is the tunnelling spin polarisation, *m* is the reduced magnetisation, $G_{\text{hop}}$ is the spin-independent conductance due to the two-step hopping via defect states in a MgO barrier. $G_0$, *m*, and $G_{\text{hop}}$ are expressed as

$$m(T) = 1 - AT^{\frac{3}{2}},$$

$$G_{\text{hop}}(T) = BT^{\frac{4}{3}},$$

$$G_0(T) = G_0 \frac{CT}{\sin CT},$$

where *A* is the parameter characterising spin fluctuation at the interface between a ferromagnet and a barrier. $C = 1.387 \times 10^{-3}\, d/\sqrt{\phi}$ with the barrier width *d* in nm and height *ϕ* in eV. *B* and $G_0$ are constants. Here, the spin-dependent and -independent parts of the conductance, $\bar{G}$ and $\Delta G$, respectively, can be defined as follows:

$$\bar{G} = \frac{G_p + G_{ap}}{2} = G_0(T) + G_{\text{hop}}(T), \tag{2}$$

$$\Delta G = \frac{G_p - G_{ap}}{2} = G_0(T)P_0^2 m^2(T). \tag{3}$$

We find Equations (2) and (3) fit the experimental data very well with *C* = 1.4 × 10[-3] as shown in Fig. 2. $G_p$, $G_{ap}$ and the TMR ratio are also found to satisfy the following relationship:

$$\text{TMR ratio} = \frac{2\Delta G}{G_{ap}} \times 100,$$

which is in agreement with the model as seen in Fig. 1(d). From this analysis, we obtained $P_0$ of 0.87 and *A* of 1.7 ×10[-5] K[-3/2]. *A* is found to be twice as large as that observed in Fe/MgO/Fe MTJs at (7-9) × 10[-6] K[-3/2] [27]. $G_{\text{hop}}$ are also plotted in the figure, which is relatively large at RT. Thus, the temperature dependence of the TMR ratios is mainly induced by the spin fluctuation at the CoMn/MgO interfaces and the spin-independent hopping within the MgO barrier.

In order to clarify the possible interfacial spin fluctuation and spin-independent hopping in the barrier, atomic structural imaging was performed on these CoMn-based MTJs, namely those for $x$ = 75 and 86 with the TMR ratios of 229 and 142%, respectively. Figures 2(a) shows a cross-sectional HRTEM image of the bcc $Co_{75}Mn_{25}$ MTJ sample with the larger TMR ratio of 229% at RT [24]. Figure 2(a) confirms the fully epitaxial growth of the entire MTJ with some dislocations. Lattice dislocation has been reported as one of the crucial factors on MTJ performance. The TMR ratio is strongly correlate to the density of lattice dislocation at the electrode/barrier interface [10][28]. Therefore, an inverse Fourier transform is carried out on the corresponding TEM images to identify lattice dislocations. The dislocation at the CoMn/MgO/CoMn interface is labelled using a red arrow as shown in Fig. 2(b). The period of the lattice dislocation is calculated to be (11.4 ± 0.3) nm. It should be noted that the dislocations also appear within the top and bottom CoMn layers almost every 1 nm along plane normal, suggesting the metastable CoMn grains have a critical size to maintain their lattice matching with the layers underneath. This indicates that the $Co_{75}Mn_{25}$ layer may induce plastic deformation rather than elastic deformation with inducing dislocations at the boundaries between crystals.

In the MgO layer, only two distinctive crystallographic phases are observed as shown as the regions (i) and (ii). Namely for the regions (i), similar crystalline structures can be found in the other reports where some have claimed that it is an amorphous structure [17][29] but the others have described it is crystallised in the (200) orientation [30]. In order to clarify the regions (i) and (ii), two inverse Fourier transformations which correspond to the MgO[001] and [100] orientations are extracted from Fig. 2(a) as indicated in Figs. 2(b) and (c), respectively. Firstly, the region (ii) shown in Fig. 2(a) is fully crystallised MgO, which will be discussed further later using the thicker MgO reference in Fig. 3. No fringe splitting is appeared within the green circle in both Figs. 2(b) and (c). The purple circle in Fig. 2(b) shows multiple fringe splitting but not in Fig. 2(c). It indicated the MgO lattice is crystallised but stretched along the [001] orientation introducing the dislocation along the [100] axis. Hence, MTJ with the larger TMR ratio is found to be fully crystallised with some lattice stretching in MgO and interfacial dislocations.

Figure 2(d) shows a cross-sectional HRTEM image of the $Co_{86}Mn_{14}$ MTJ sample with the smaller TMR ratio of 142% at RT. The boundary between CoMn[110] and MgO[100] layers can be clearly observed in this image. In Fig. 2(d), the dark black regions at the CoMn/MgO interfaces indicate where CoMn/MgO is not fully crystallised. Such features are typically observed at the CoMn/MgO interface and within the CoMn layer. These features are found to be induced by dislocations formed at the CoMn/MgO/CoMn interface as labelled using a red arrow in Fig. 2(e). One of the interesting features is that the dislocations often appear near the dark black regions which are circled in red in Fig. 2(d). This pattern can also be observed in the earlier reports, but no specific comments were made to date [9]. In this study, we demonstrate the appearance of fringe splitting at the CoMn/MgO/CoMn interfaces is not only due to the lattice dislocation but a partially crystallised structure. The period of the lattice dislocation is calculated to be (8.9 ± 0.3) nm, which is shorter than that for the larger TMR sample. These results coincide with the correlation between the interfacial dislocation density and the TMR ratio correlation reported by Bonell *et al.* [31] and Liu *et al.* [32]. These dislocations and the associated interfacial mixing/disordering may be the origin of the interfacial spin fluctuation as identified in Fig. 1, *i.e.*, large $A$. Note that the dislocations are also formed within the top and bottom CoMn layers as the same with the $x$ = 75 sample as discussed above. However, their density in the top CoMn layer for $x$ = 86 is much higher than that for $x$ = 75, while that in the bottom is almost the same. This may be due to the larger discrepancy in the lattice constants between CoMn and MgO as discussed later.

In the MgO barrier, two regions (i) and (iii) can be unambiguously distinguished in Fig. 2(d) as fully crystallised and partially crystallised MgO, respectively. The grain size of these regions is measured to be 2 ~ 4 nm in diameter. Such inhomogeneous multi-boundaries in the MgO barrier layer indicate the MgO layer is not epitaxially grown along the [100] orientation. Figure 2(e) is the corresponding inverse Fourier transform corresponding to the MgO[001] of Fig. 2(d). The fringe splitting in Fig. 2(e) is observed within the MgO as labelled as a yellow circle, which is the region where the partially crystallised MgO was identified. Whereas the green circle labelled in Fig. 2(a) is the region where no fringe splitting was observed as represented as crystallised MgO in the region (iii). The lattice mismatch is possible to appear within the layer itself due to local stress as seen in Figs. 2(b) and (c) or partial crystallisation as seen in Fig. 2(e). It can hence be concluded that the fringe splitting appears within the layer is due to the partially crystallisation. Whilst the fringe splitting appears at the interface is caused either by the lattice dislocation or partially crystallised structure.

In order to achieve a high TMR ratio, high (or full) crystalline ordering is required since it improves the band matching between the electrodes and barrier for the coherent tunnelling [10, 31]. Hence, we assume the reduction in the TMR ratio originates from local lattice defects at the interface, electrodes and barrier layers appeared at the boundaries of partially crystallised regions.

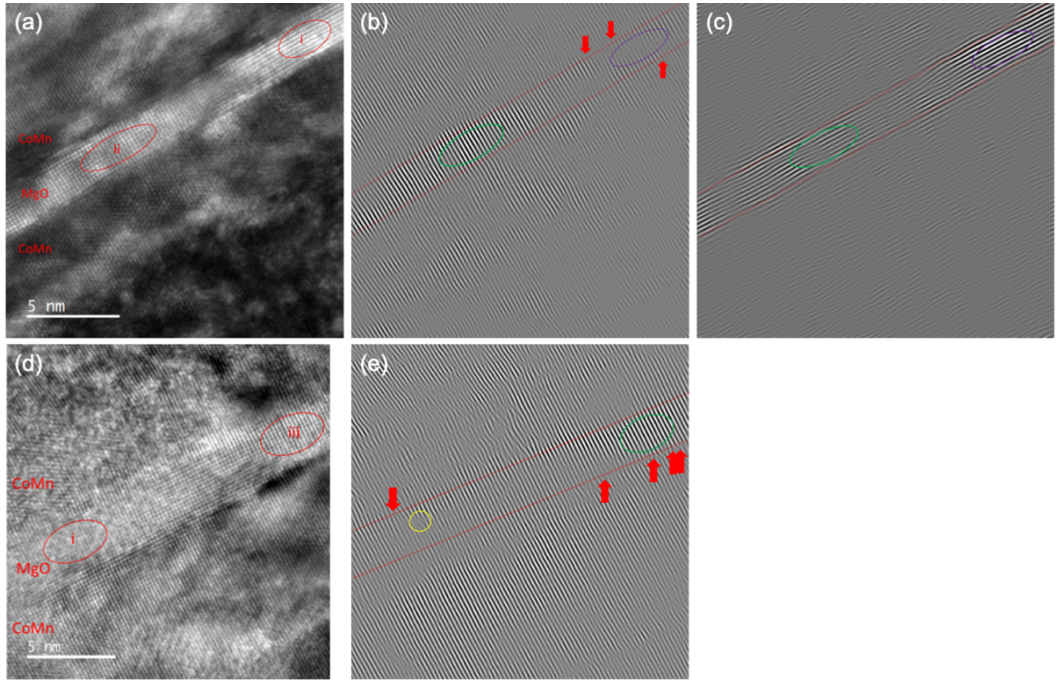

Figure 2 HRTEM images taken on the (a) $Co_{75}Mn_{25}$ [24] and (d) $Co_{86}Mn_{14}$ samples. The corresponding Fourier transformations are shown in (b) along the MgO[001], (c) [100] and (e) [001].

The crystallisation degrees of the MgO barrier are further studied by using reference samples with the thicker MgO barrier, consisting of MgO substrate/Cr (40)/$Co_xMn_{100-x}$ (10)/MgO (20)/$Co_xMn_{100-x}$ (4)/CoFe (1.5)/IrMn (10)/Ru (5) (thickness in nm, $x$ = 75). The MgO layers in the reference samples are approximately 10 times thicker than previous set of MTJ samples to obtain clear selected area electron diffraction (SAED) patterns, allowing to investigate further MgO crystallographic structures. The cross-sectional TEM specimens were prepared using focused ion beam (FIB, Nova 200 NanoLab). The samples were first milled down to a thickness of < 100 nm and were polished down to electron transparency by a Gatan gentle mill (Scientific Technical Development, GENTLE MILL ion miller model IV5).

In Fig. 3, a cross-sectional HRTEM on $Co_{75}Mn_{25}$ (10)/MgO (20)/$Co_{75}Mn_{25}$ (4) (thickness in nm) sample with the corresponding SAED image. Here, three different regions of the 20 nm thick MgO layers are analysed over the TEM images to estimate their volume ratios: (a) approximately 58% of crystallised MgO[100] as labelled as the regions (i) and (ii) in Fig. 2, (b) ~ 4% of crystallised MgO with different orientations as seen as the region (iii) in Fig. 2 and (c) almost 38% of partially crystallised MgO. The partially crystallised MgO only appears in the centre of the MgO barrier, which is due to the increased thickness of the MgO barrier, possibly indicating that the MgO crystallisation is initiated at the CoMn/MgO interfaces and is not reached to the centre under the post-annealing condition used. Figure 3(a) shows crystallised MgO where highly ordered electron beam diffraction spots can be observed in selective diffraction patterns. The red squares in Fig. 3(a) show the region where MgO is crystallised along the [100] direction. A similar crystalline structure is observed in Fig. 2(e) but with an additional phase with a different crystalline direction. The blue circles shown in the diffraction pattern in Fig. 3(b) indicate missing diffraction spots, which is a sign of lower degree of crystallinity than that in Fig. 3(a). Nevertheless, the first order diffraction spots are identical to those in Fig. 2(d), indicating that a part of the MgO layer is orientated in a different direction with respect to the [100] axis. Furthermore, the diffraction pattern in Fig. 3(c) represents the dark grey region in the selected MgO layer. Similarly, some of the second order diffraction spots are vanished. The unique feature in this figure is that the first and second order diffraction spots are rather dispersive, which apparently confirms that MgO is not fully crystallised as discussed above. Such partially crystallised MgO can be the origin of the spin-independent hopping and the grain boundaries between the MgO grains with three different phases can act as hopping sites.

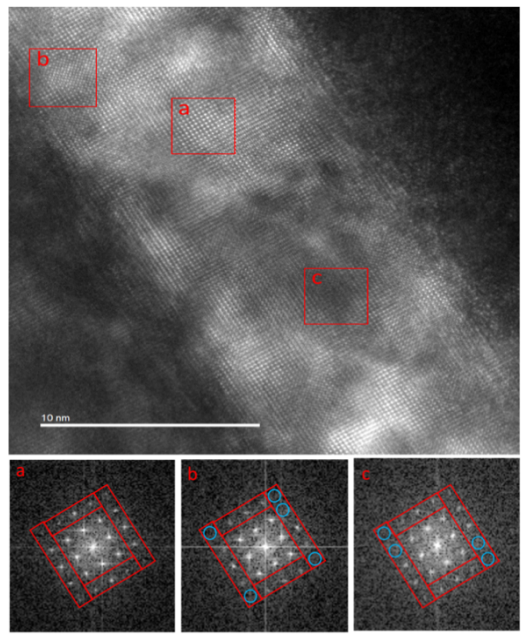

Figure 3 HRTEM image of the reference sample, MgO substrate/Cr (40)/Co$_{75}$Mn$_{25}$ (10)/MgO (20)/Co$_{75}$Mn$_{25}$ (4)/CoFe (1.5)/IrMn (10)/Ru (5) (thickness in nm), focusing on the MgO layer with three distinctive regions confirmed by the diffraction patterns: (a) MgO[100]; (b) MgO with a different orientation from MgO[100] and (c) partially crystallised MgO.

Since MTJ is constructed with two CoMn layers, the lattice constants of the Co$_x$Mn$_{100-x}$ layers are measured and compared to investigate their softness. The lattice constants are calculated by measuring the averaged fringe distance across 20 layers in the corresponding TEM images. Figure 4 shows the lattice constants of the top and bottom CoMn layers for MTJs with four different CoMn compositions. It shows that apart from the $Co_{86}Mn_{14}$ MTJ, the top CoMn lattice constant is about 3.5% larger than the bottom layer across MTJs. The largest difference is measured to be 4.3% in the $Co_{75}Mn_{25}$ MTJ. The enhancement in the CoMn lattice constants is mainly induced by the material underneath the CoMn layer. The measured values of the Cr and MgO lattice constant are 0.27nm and 0.40nm, respectively. This is comparable to the theoretical value of 0.29nm and 0.42nm [28]. Assuming the Cr/CoMn and CoMn/MgO interface have 45º ($\times \sqrt{2}$) rotation, the lattice constant of Cr became 0.41 nm. Therefore, it consists of about 2.4% difference in the lattice constants [9], which agrees well with the measured values. As the bottom CoMn layer is grown on the Cr seed layer, while the top CoMn is grown on the MgO barrier, the CoMn lattice constants may vary due to the different texture of the underlayers.

A monotonic increase in the CoMn lattice constants for the CoMn films annealed at 200C has been reported by X-ray diffraction (XRD) with increasing the Mn contents エラー! 参照元が見つかりません。. As XRD is a macroscopic measurement, the estimated lattice constants are averaged values over the bottom CoMn layers. Kunimatsu *et al.* reported that the lattice constants along the plane normal direction show very minor increase from 0.285 nm at $x$ = 86 to 0.290 at $x$ = 50 almost linearly, while the lattice in the in-plane [001] direction stays almost constant (~ 0.288 nm). This trend generally agrees with that of the top and bottom CoMn layers measured in the TEM images as shown in Fig. 4, except for $x$=0.86. This may be due to the CoMn lattice strain induced by the MgO tunnel barrier and the Cr seed layer underneath, especially for the top CoMn layer for $x \leq 83$. Note that the number of dislocations almost stays the same for both top and bottom interfaces as seen in Fig. 2(a), confirming the soft nature of the lattice constants of CoMn and MgO as appeared in Fig. 4.

For $x$ = 86, the bottom CoMn lattices is found to increase, introducing a large number of dislocations at the bottom CoMn/MgO interface as seen in Fig. 2(d). The top CoMn layer, on the other hand, absorbs the strain by introducing more dislocations within the layer due to the larger difference in the lattice constants between CoMn and MgO underneath.

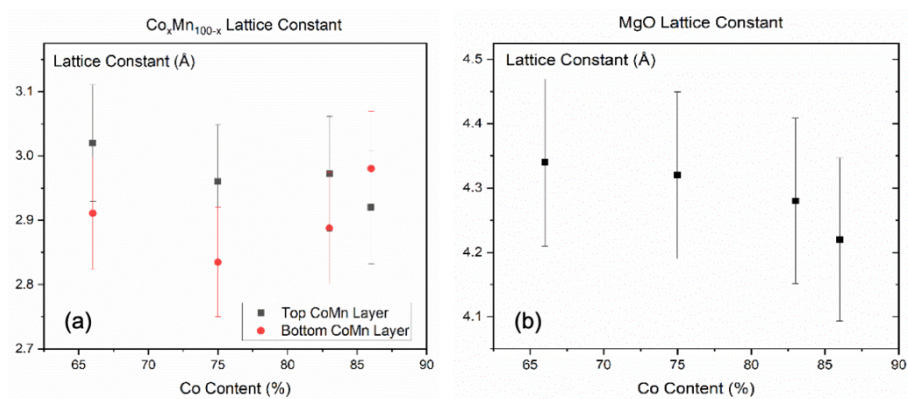

Figure 4 (a) Lattice constants of $Co_xMn_{100-x}$ layers which are located at the top and bottom of the MgO layer with different compositions. (b) Lattice constant of MgO on different $Co_xMn_{100-x}$ compositions.

The TMR ratios of $Co_{75}Mn_{25}$ and $Co_{86}Mn_{14}$ were measured to be 229% and 142%, respectively. The TMR ratio is dependent on several factors, including the crystallinity of MTJs, potential barrier height and thickness, interfacial roughness and bonding at the electrode insulator interface [10]. In this study, the TMR ratio is mainly analysed in terms of the crystallisation between the ferromagnetic and barrier layers. For the $Co_{75}Mn_{25}$ MTJ as shown in Fig. 2(a), the dark black regions, *e.g.*, one at the top CoMn/MgO interface at the left end of the image, are formed with the total length across 7 nm out of 40 nm (17% effective area). As discussed above, these regions only contain dislocations with maintaining epitaxial growth due to the softening of the CoMn layers. It is also confirmed that the MgO barrier is crystallised entirely with some stretched regions. Whilst, for the $Co_{86}Mn_{14}$ MTJ in Fig. 2(d), the dark black regions located at the CoMn/MgO interfaces appeared in with a total length of approximately 10 nm out of 40 nm (25% effective area). These regions are characterised as partially crystallised regions. Additional distributions in the MgO crystalline orientations are imaged in the barrier. Such inhomogeneous crystallisation at the CoMn/MgO interfaces and the MgO barrier layer are attributed to the reduction in TMR ratios for $x = 75$.

As reported earlier in $Fe_{70}Pd_{30}$ alloys [33], softening of a ferromagnetic lattice allows the junction to form a strain-less interface, which is critical to eliminate any unexpected electron scattering and spin fluctuation. For the metastable bcc CoMn alloys investigated here, we carried out first-principles density-functional calculations to gain the insight into the lattice deformation of metastable bcc CoMn alloys. The calculations have been performed using the projector augmented plane-wave (PAW) method [34] implemented in the Vienna *ab-initio* simulation package (VASP) code [35][36]. We adopted the generalised gradient approximation [37] for exchange and correlation energies/potentials. We used a $6 \times 6 \times 6$ ***k***-point mesh for Brillouin zone integration and 500 eV for a cut-off energy of plane-wave expansion. We constructed a 20-atom supercell and considered special quasi-random structures (SQSs) [38] to simulate chemical disorder. The SQSs were generated using the alloy theory automated toolkit package [39].

Figure 5(a) shows the total energies of ferromagnetic $Co_xMn_{100-x}$ alloys with various compositions calculated as a function of the strain, $\delta$, for volume-conserving tetragonal distortion. As shown in Fig. 5(a), the ferromagnetic bcc phase is metastable for $x \leq 90$, where the curvature of total energy around the bcc phase becomes smaller with increasing the Co composition $x$, and eventually it becomes unstable for $x \geq 95$. In Fig. 5(a), the total energy for ferromagnetic Fe is also shown for comparison. The curvature of total energy is much smaller for the ferromagnetic CoMn alloys in the Co composition range, $65 \leq x \leq 90$, compared to that for ferromagnetic bcc Fe. The result suggests that the bcc CoMn alloys are much softer against the tetragonal distortion than bcc Fe. For more quantitative comparison the tetragonal shear modulus $C'$ was evaluated from the curvature of total energy in Fig. 5(a). The results are plotted as a function of the Co composition $x$ in Fig. 5(b). The obtained values of $C'$ are 35.5 and 32.5 GPa for $Co_{65}Mn_{35}$ and $Co_{75}Mn_{25}$, respectively. These values are about one half of $C'$ evaluated for bcc Fe (61.8 GPa). Figures 5(a) and (b) show the softening of bcc $Co_xMn_{100-x}$ becomes prominent near the boundary of the metastable bcc phase. Although the maximum softness is expected for $x = 90$, it is difficult to stabilise a bcc CoMn phase in the form of MTJ as discussed above. This may leave some additional potential of this alloy for further improvement in the corresponding magnetotransport properties.

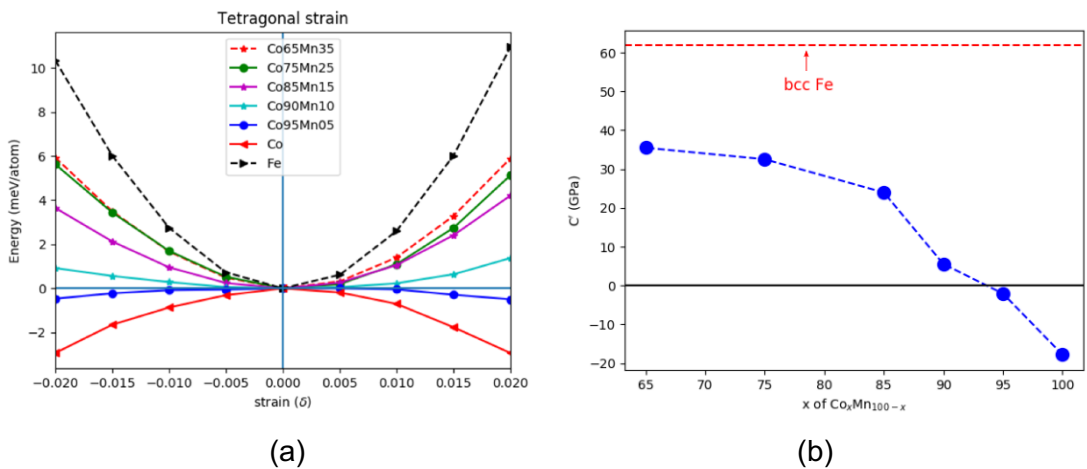

(a)          (b)

Figure 5 (a) Total energies of ferromagnetic $Co_xMn_{100-x}$ alloys with $x$ being 65, 75, 85, 90, 95 and 100 calculated as a function of the strain, $\delta$, for tetragonal distortion. The bcc structure corresponds to $\delta = 0$. The total energy of ferromagnetic Fe is also shown for comparison. (b) Tetragonal shear moduli $C'$ of ferromagnetic $Co_xMn_{100-x}$ alloys as a function of the Co composition $x$. A horizontal broken line represents the value of $C'$ for bcc Fe.

    The use of a soft ferromagnetic layer can offer a similar effect as the precise lattice matching of a tunnel barrier with the neighbouring ferromagnets. Belmoubarik *et al.* have developed a spinel barrier, of which lattice constant can be precisely controlled by its composition to match with that of the neighbouring ferromagnetic layers [40], which removes the interfacial spin fluctuation and spin-independent hopping in the barrier. They have fabricated an epitaxial MTJ consisting of $Fe/MgAl_2O_4/Fe(100)$, demonstrating TMR ratios of

245 and 436% at RT and 3K, respectively. They have further improved the TMR ratios up to 342% at RT by replacing Fe with a $Co_2FeAl$ Heusler alloy with large spin polarisation [41]. These TMR ratios are almost the same with our case. Assuming the spin polarisations of Fe and CoMn are comparable, the MgO barriers may be able to maintain coherent tunnelling even under the presence of the stretched regions in the barrier, which may possibly induce the spin-independent hopping. This is advantageous over the other tunnel barriers to maintain large TMR ratios with some distributions in their crystalline orientations. Hence, the MgO barriers holds a space for further improvement in the TMR ratios closer to the theoretically predicted values by suppressing the interfacial spin fluctuation.

**Conclusion**

Our investigation confirms the softening of metastable bcc $Co_xMn_{100-x}$ ferromagnetic layers for $x ≤ 83$, achieving the large TMR ratio of 229% at RT, which is comparable with the strain-free MTJ with a spinel barrier. These MTJs show less dislocations at the CoMn/MgO interfaces due to the CoMn softening and MgO stretching. This proves the robustness of MgO for coherent tunnelling. For $x = 86$, on the other hand, due to the MgO contraction, the higher density of the lattice dislocations is observed, one possible source in the lower TMR ratio of 142%. The HRTEM images additionally show that the MgO barrier forms fully crystallised along not only the [100] direction but also non-[100] directions for $x = 86$. This inhomogeneous crystallisation partially induced by the lattice mismatching causes the suppression of electron tunnelling through the barrier and leads to the reduction in the corresponding TMR ratio. The first-principles calculation also confirms the softening of bcc $Co_xMn_{100-x}$, which becomes prominent near the boundary of the metastable bcc phase. Although the maximum softness is expected for $x = 90$, it is difficult to stabilise the bcc CoMn phase in the form of MTJ after post-annealing as experimentally observed for the case of $x = 86$. Further precise control of the lattice matching can demonstrate much larger TMR ratios as required for greater integration and capacity of MTJ-based storages and memories.

**Acknowledgments**

This work was partially supported by JST CREST (No. JPMJCR17J5).